\documentclass[aps,prl,twocolumn,groupedaddress,showpacs,floatfix]{revtex4}
\def\bc{\begin{center}}
\def\ec{\end{center}}
\def\be{\begin{equation}}
\def\ee{\end{equation}}
\renewcommand{\vec}[1]{\mbox{\boldmath$#1$}}
\usepackage{longtable}
\usepackage{epsfig}

\begin{document}

\title{Composite-fermion crystallites in quantum dots}
\author{Gun Sang Jeon, Chia-Chen Chang, and Jainendra K. Jain}
\affiliation{Physics Department, 104 Davey Laboratory, The Pennsylvania State University,
University Park, Pennsylvania 16802}

\date{\today}

\begin{abstract}
The correlations in the ground state of interacting electrons in 
a two-dimensional quantum dot in a high magnetic field
are known to undergo a qualitative change from liquid-like to crystal-like
as the total angular momentum becomes large.
We show that the composite-fermion theory provides an excellent account 
of the states in both regimes.  The quantum mechanical formation of 
composite fermions with a large number of attached vortices automatically generates 
omposite fermion crystallites in finite quantum dots.
\end{abstract}
\maketitle

The system of interacting electrons confined to a two dimensional 
quantum dot and exposed to a strong magnetic field has been a subject 
of intense theoretical study for over two decades~\cite{Yoshioka,Girvin,
Dev,Yang,Tapash,Xie,Hawrylak,Kawamura,Kamilla,Seki,Manninen,Landman,Ruan,Maksym}, 
both because such quantum dots have been realized and studied in 
the laboratory~\cite{Expt1,Expt2,Expt3}, and because of the 
possible relevance of its physics 
to the fractional quantum Hall effect (FQHE)~\cite{Tsui}.
For systems with small numbers of particles, exact diagonalization 
can be performed for parabolic quantum dots in the limit when the 
cyclotron energy is large compared to the confinement potential,
which shows that the ground state energy as 
a function of the angular momentum ($L$) has 
a rather rich structure.
In particular, downward cusps appear at certain values of $L$, 
which are consequently especially favorable.  An understanding of 
the correlations in the quantum dot state that underlie  
this physics is one of the central questions for this system.  
One would also like to know how this 
ties into our understanding of the FQHE, obtained in the 
thermodynamic limit without confinement.

The approach based on the formation of composite fermions 
has been demonstrated to be successful  
in a range of $L$ values~\cite{Kawamura,Kamilla}.  Specifically, 
a mean-field type description, in which the composite fermions
are taken as non-interacting particles at an effective angular momentum
$L^*$, with their mass or the cyclotron energy
treated as a phenomenological parameter, predicts cusps
in the energy at certain magic angular momenta, which are in
agreement with the actual cusp positions in exact 
diagonalization studies~\cite{Kawamura}.  However, discrepancies 
appear at large $L$~\cite{Seki,Landman}; here,
the actual cusps occur at regular intervals in $L$, 
which has been interpreted in terms of classical crystal-like 
states~\cite{Ruan}.  At large $L$, the repulsive interaction thus 
appears to stabilize a (rotating) crystal 
rather than a liquid, which is believed to signal 
a breakdown of the composite-fermion (CF) description.  That would 
not be unexpected, because large angular momenta correspond to small 
filling factors (to the extent the 
filling factor is a meaningful quantity in a finite quantum dot),
and it is known, for infinite two-dimensional systems, 
that the CF liquid gives way to a Wigner crystal at 
sufficiently small fillings.

To understand the nature of the 
breakdown of the composite fermion description, a matter 
of great interest also in the context of the FQHE, we 
have undertaken an extensive study of finite systems at large angular 
momenta.  Our investigations, however, have led to a surprising conclusion:
Even though the naive {\em mean-field} interpretation in terms of {\em free} 
composite fermions becomes invalid at large $L$,
the microscopic composite fermion theory, defined through 
wave functions, continues to give a very good description down 
to the largest $L$ studied to date.   
It provides an accurate  approximation for the ground state wave function 
and the ground state energy at every single $L$ in the 
wide range studied, and correctly reproduces all cusps 
in plot of the ground state energy vs. $L$.  Taken together, these 
results constitute a detailed verification for  
the validity of the composite fermion theory for quantum dots even 
at very low fillings.

\begin{figure}
\centerline{\epsfig{file=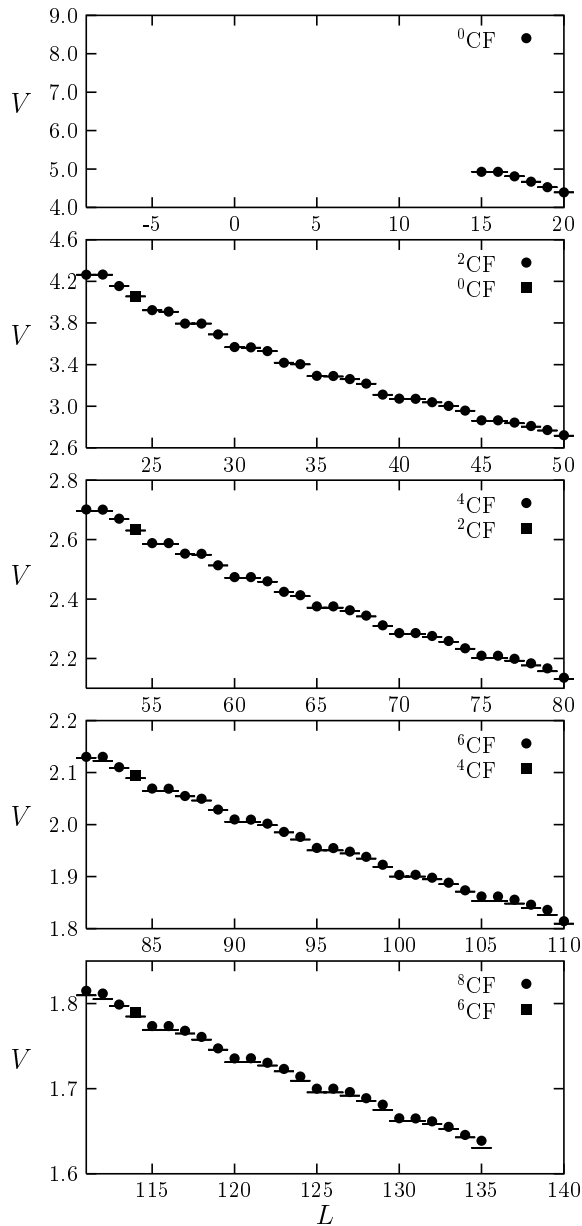,width=3.0in,angle=0}}
\caption{
The exact interaction energy $V$ (dashes) is given as a function of the angular momentum
$L$ for $N=6$ particles.  The dots are the predictions of the CF theory.
Different panels correspond to $L$ regions where composite fermions of 
different flavors are relevant. (In each panel, for $L$ related to 
$L^*=-6$, the lower order CF gives a smaller $D^*$.)
The energies are quoted in units of $e^2/\epsilon \ell$.
}
\label{fig:energy}
\end{figure}

This may appear to be at odds with the 
classical crystal-like correlations found in exact diagonalization 
studies~\cite{Seki,Maksym}.  While both composite fermions and the 
crystal are generated by the repulsive interaction between electrons, 
the feeling has been that one excludes the other.  
A notable aspect of our work is the unexpected finding that 
not only is there no logical inconsistency between the simultaneous 
formations of composite fermions and crystal-like structures at 
low fillings, but that the state is very well described as 
a crystallite of composite fermions.

The Hamiltonian of interest is  
\be
H=\sum_j\frac{1}{2m_b}\left(\vec{p}_j+\frac{e}{c}\vec{A}_j\right)^2
+\sum_j\frac{m_b}{2} \omega_0^2 r_j^2 + 
\sum_{j<k} \frac{e^2}{\epsilon r_{jk}}
\ee
where $m_b$ is the band mass of the electron, $\omega_0$ 
is a measure of the strength of the confinement, $\epsilon$ is 
the dielectric constant of the host semiconductor, and $r_{jk}=
|\vec{r}_j-\vec{r}_k|$. 
We will specialize to the case of very large magnetic fields,
when $\omega_c = eB/m_b c >> \omega_0$.  Only the lowest Landau level
(LL) is relevant in this limit. 
In that limit, at each angular momentum the 
eigenenergy neatly separates into a confinement part and an 
interaction part:
\be
E(L)=E_c(L) + V(L)
\ee
where 
$E_{c}(L)=(\hbar/2)[\Omega-\omega_{c}] L$,
relative to the lowest LL, with 
$\Omega^2=\omega_{c}^2+4\omega_{0}^2$,
and $V(L)$ is the interaction energy of electrons without 
confinement, but with the magnetic length replaced by 
an effective magnetic length given by $\ell\equiv
\sqrt{\hbar/m_b\Omega}$. 
In the following, we will only consider $V$ as a function of the 
angular momentum $L$; it must be remembered, however, 
that the confinement part must eventually be added to determine the 
global ground state. 
The exact ground state in each $L$ sector is obtained by 
either matrix diagonalization or the Lanczos method.
The largest system we have studied has a Fock space dimension of 509,267.

In the CF theory~\cite{Dev,Kawamura,Jain}, the interacting state of electrons 
in the lowest LL at angular momentum $L$ is 
mapped into the non-interacting electron state at $L^*$,
where
\begin{equation}
L=L^*+pN(N-1)\;,
\end{equation}
$N$ is the number of electrons, and $p$ is an integer.  
The wave functions are related as
\begin{equation}
\Psi^L_\alpha={\cal P} \prod_{j<k}(z_j-z_k)^{2p} \Phi^{L^*}_\alpha .
\end{equation}
Here $\Phi^{L^*}_\alpha$ are the wave functions for non-interacting 
electrons at $L^*$ (which in general occupy 
several Landau levels), $\alpha=1, 2, \cdots, D^*$ labels the different states,
$z_j=x_j-iy_j$ denotes the position of the $j$th electron, 
${\cal P}$ indicates projection into the lowest LL, 
$\Psi_\alpha$ are basis functions for interacting electrons 
at $L$, and $D^*$ is the dimension of the 
CF basis.  We will restrict $\Phi_\alpha$ to states with the lowest
kinetic energy at $L^*$, and choose $p$ so as to have the smallest 
dimension for the basis.  The composite fermions carrying $2p$ vortices
are labeled $^{2p}$CF's.
At certain values of $L$, the above prescription produces only one 
state ($D^*=1$), which is the CF theory's answer for the 
ground state.  In the notation of 
Ref.~\cite{Kawamura}, this is a compact state, denoted by $(N_0,N_1,\cdots)$,
with $N_j$ composite fermions compactly occupying the innermost angular momentum
orbitals of the $j$th CF level.  At other values of $L$, 
when there are many CF basis states ($D^*>1$), we diagonalize the 
Coulomb Hamiltonian in the CF basis to obtain the ground state,
using methods described earlier~\cite{Kamilla,Mandal}.
For any $N$, there are many values of $L$ where the CF theory 
gives a unique answer, but in general, $D^*$ increases with $N$.

\begin{figure}
\centerline{\epsfig{file=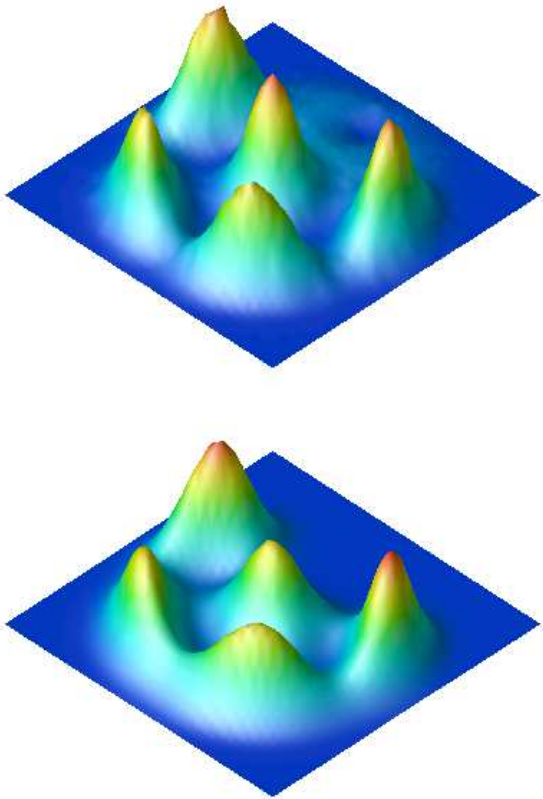,width=3.0in,angle=0}}
\caption{The upper plot shows the pair 
correlation functions for the exact 
ground state for $N=6$ particles at $L=95$, and the lower plot displays the 
prediction of the CF theory.  The exact ground
state is a linear superposition of 69,624 basis states, whereas the 
$^6$CF ground state is given by 
a unique wave function denoted by $(4,2)$.  The ``missing
peak" on the outer ring indicates the location $\vec{R}$ of the fixed particle. 
}
\label{fig:pairc}
\end{figure}

We have carried out exact diagonalization for $N\leq 8$. 
It is possible to to go to arbitrarily large values of $L$
within the CF theory, but available computer memory has restricted our 
{\em exact} diagonalization study to $L \leq 135, 117$ and 111 
for $N=6$, 7 and 8, respectively. 
Figure~\ref{fig:energy} shows the comparison between the CF theory 
and the exact energy as a function of $L$ for $N=6$ particles,
demonstrating that the CF theory predicts every energy accurately
and obtains all cusps faithfully \cite{Comment}.  The comparisons 
for other values of $N$ are similar.

We next come to 
the connection between our approach and the crystalline correlations
found in previous studies.  Such correlations are not manifest in 
the density, which is rotationally symmetric for the ground state
with a definite angular momentum,
so one must compute the pair correlation function
\begin{equation}
g(\vec{r})\sim \int \prod_{j=3}^N d^2 \vec{r}_j 
|\Psi(\vec{r},\vec{R},\vec{r}_3,\cdots, \vec{r}_N)|^2
\end{equation}
which is the probability of finding a particle at $\vec{r}$
while holding one particle fixed at $\vec{R}$.  We first 
compute the density as a function of the distance from 
the center, and fix $\vec{R}$ somewhere on the maximum density
ring.  As an example, Fig.~\ref{fig:pairc} (upper panel) shows the 
exact pair correlation function for 6 particles at $L=95$, which 
clearly illustrates crystallite formation. The CF theory gives 
a unique state here, $(4,2)$, which has the wave function 
(with $N=6$ and $2p=6$) 
\begin{equation}
\Psi = {\cal P} A\left[ z_{1}^*\cdot z_{2} z_{2}^*\cdot  
\prod_{i=3}^{N}  z_i^{i-3} \right]
\prod_{j<k}(z_j-z_k)^{2p} e^{-\sum_{l=1}^N |z_l|^2/4}
\label{wf}
\end{equation}
where $A$ denotes an antisymmetric Slater determinant.
The projection is accomplished by the method outlined in the 
literature~\cite{Kamilla}.
The energy of this wave function is $1.95535(15) e^2/\epsilon\ell$,
which compares well with the exact energy $1.95061 e^2/\epsilon\ell$. 
The overlap with the exact ground state is 0.902.
The pair correlation function for this wave function 
is also shown in Fig.~\ref{fig:pairc} (lower plot).
It is remarkable that the single wave
function from the CF theory provides a good
qualitative and quantitative description of the actual 
ground state wave function in Fig.~\ref{fig:pairc} 
which is a linear combination of 69,624 basis functions.
We have studied other ``cusp states" and 
found similarly good agreement.

\begin{figure}
\centerline{\epsfig{file=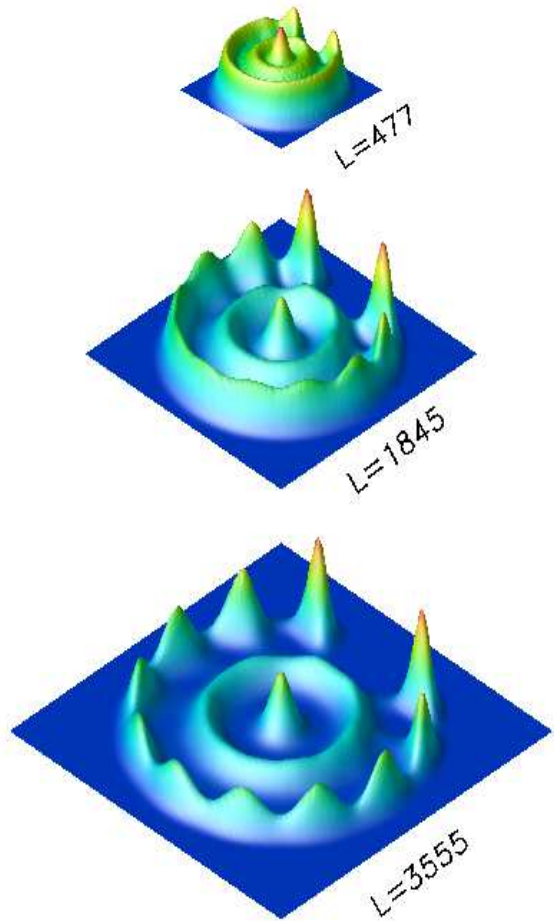,width=3.0in,angle=0}}
\caption{
Pair correlation functions for the $(17,2)$ configuration for $N=19$ 
composite fermions of three different flavors, namely 
$^2$CF's, $^{10}$CF's, and $^{20}$CF's (from top to bottom).  The missing
peak on the outer ring indicates the location $\vec{R}$ of the fixed particle. 
The three figures are plotted on the same length scale to  
illustrate how the size grows with $L$. 
}
\label{fig:pairc19}
\end{figure}

The composite fermion wave function at any angular momentum $L$ 
is obtained from {\em non}-interacting electron wave function
at an effective  angular momentum $L^*$ by multiplication by 
an appropriate Jastrow factor, in exactly the same way as 
wave functions are written for bulk fractional quantum Hall states.  
It is crucial to note that no physics of any kind of crystal is put 
into these wave functions by hand at any stage.  For example, 
referring to Fig.~\ref{fig:pairc}, the parent electron wave function 
at $L^*=5$, namely $A[\cdots]$ in Eq.~(\ref{wf}), has no crystalline correlations 
in it.  It describes a simple {\em non-interacting} system with single particle 
orbitals $\{0,0\}$, $\{0,1\}$, $\{0,2\}$, $\{0,3\}$, $\{1,-1\}$, and
$\{1,0\}$ occupied, where $\{n,m\}$ denotes the angular momentum 
$m$ orbital in the $n$th LL.  (The allowed values are $n=0,1,\cdots$, 
and $m=-n,-n+1, \cdots$.)  However, when the 
uncorrelated wave function at $L^*$ is multiplied by an appropriate 
Jastrow factor to convert electrons into composite fermions, 
the resulting wave function describes a correlated electron liquid 
for a certain range of $L$ values, but for sufficiently large angular 
momenta, it produces a state with crystalline correlation; 
in other words, the formation of composite fermions automatically causes the 
formation of a crystallite at large $L$.  It is stressed that the same 
class of wave functions 
describes both the liquid and the crystal in quantum dots in different 
parameter regimes.  The emergence of such classical crystal-like structures 
in the intrinsically quantum mechanical state of composite fermions 
is striking.

To investigate the evolution with the 
number of attached vortices, we show in Fig.~\ref{fig:pairc19} 
the pair correlation function for $N=19$ composite fermions in the 
state $(17,2)$ (with wave function given in Eq.~\ref{wf})
for three flavors of composite fermions, with $2p=2$, 10, and 20.  
The crystal becomes better defined with 
increasing $2p$.  In this case, we do not have exact results to 
compare with (the dimension of the Fock space is astronomically large; 
even for the smallest system in Fig.~\ref{fig:pairc19}, namely
19 particles at $L=477$, it is 
estimated to be $10^{14}$), but the structure at large $L$ is consistent with 
classical considerations that predict a 
three-ring structure with 1, 6 and 12 electrons located on
inner, middle and outer rings, respectively~\cite{classical}.  
Interestingly, when we hold a particle in the middle ring fixed, the
hexagonal correlations of the middle ring become prominent while the
outer ring becomes more or less uniform.  That is consistent with 
earlier studies that show that the radial correlations 
between different rings are weaker than the angular correlations 
within a ring, producing, for example, two different melting 
transitions \cite{melting}.

A recent article by Yannouleas and Landman \cite{Landman}
has asserted that the CF theory fails to produce the cusp positions 
for large $L$, especially beyond $L>75$ (for $N=6$).
That assertion is not borne out by our results, which
demonstrate that the CF theory, when taken beyond the most 
naive mean field picture, predicts each cusp correctly at least for
angular momenta up to $L=135$.  Ref.~\onlinecite{Landman}
has studied an alternative trial
wave function based on an analogy to the classical crystal
ground state in a quantum dot.  A thorough comparison of the
CF theory and the approach of Ref.~\onlinecite{Landman} is
not possible because the latter obtains wave functions and
energies only for certain special values of $L$.
For $N=6$, Ref.~\onlinecite{Landman} explicitly quotes 
energies from their approach for seven values of $L$ in the 
range $75\leq L \leq 135$.  For these seven angular momenta, 
the CF theory gives lower energy in every case except 
at $L=135$.  In future, it would be interesting to compare 
the two methods for a larger range of $L$ and $N$ to ascertain 
their respective regimes of validity.

One may ask what implications our results have for the nature of the 
bulk FQHE state at small fillings.  While the finiteness of 
the systems does not allow us to draw a firm conclusion regarding
whether the thermodynamic state at some filling is a liquid or
a crystal, the results do indicate that the CF state 
possesses a substantial short-range crystalline order at small 
$\nu$.  This raises the interesting issue of whether the actual
Wigner crystal at low fillings is a crystal
of electrons or of composite fermions~\cite{Fertig}.

The composite fermion description of correlated electron states in 
quantum dots at high magnetic fields has several appealing features:
It gives a unified theory, applicable to angular momenta 
spanning {\em both} the 
liquid and the crystal phases; it produces microscopic wave functions 
also for those angular momenta where no classically stable crystal is 
available; it obtains the correct crystal shape and the ring structure 
without an explicit consideration of the classical solution; 
and finally, through CF theory, the understanding of the quantum dot physics  
dovetails nicely with that of the FQHE.  
Before ending, we note that the accuracy of the CF wave functions can be 
improved straightforwardly and systematically 
by enlarging the basis at $L^*$ slightly~\cite{Peterson}; the goal 
in this work was to demonstrate that the zeroth order theory itself works 
very well.

Partial support by the National Science Foundation under grant
no. DMR-0240458 is gratefully acknowledged.


\begin{thebibliography}{99}

\bibitem{Yoshioka} D. Yoshioka, B.I. Halperin, and P.A. Lee, 
Phys. Rev. Lett. {\bf 50}, 1219 (1983).

\bibitem{Girvin} S.M. Girvin and T. Jach, Phys. Rev. B {\bf 28}, 4506 
(1983).

\bibitem{Dev} G. Dev and J.K. Jain, Phys. Rev. B {\bf 45}, 1223
(1992).

\bibitem{Yang} S.-R. Eric Yang, A.H. MacDonald, and M.D. Johnson, Phys.
Rev. Lett. {\bf 71}, 3194 (1993).

\bibitem{Tapash} P.A. Maksym and T. Chakraborty, Phys. Rev. Lett. {\bf
65}, 108 (1990).

\bibitem{Xie} X.C. Xie, S. Das Sarma, and S. He, Phys. Rev. B
 {\bf 47}, 15942 (1993).

\bibitem{Hawrylak} P. Hawrylak, Phys. Rev. Lett. {\bf 71}, 3347 (1993).

\bibitem{Kawamura} J.K. Jain and T. Kawamura, Europhys. Lett. {\bf
	29}, 321 (1995).

\bibitem{Kamilla} J.K. Jain and R.K. Kamilla, Int. J. Mod. Phys. {\bf
	11}, 2621 (1997). 

\bibitem{Seki} T. Seki, Y. Kuramoto, and T. Nishino, J. Phys. Soc.
Jpn. {\bf 65}, 3945 (1996).

\bibitem{Manninen} M. Manninen, S. Viefers, M. Koskinen, and S. M. Reimann, 
Phys. Rev. B {\bf 64}, 245322 (2001). 

\bibitem{Landman} C. Yannouleas and U. Landman, Phys. Rev. B {\bf 68}, 
035326 (2003).

\bibitem{Ruan} W.Y. Ruan, Y.Y. Liu, C.G. Bao, and Z.Q. Zhang, Phys.
Rev. B {\bf 51}, R7942 (1995);
W.Y. Ruan and H.-F. Cheung, J. Phys.: Condens. Matter {\bf 11}, 435
(1999).

\bibitem{Maksym} P.A. Maksym, Phys. Rev. B {\bf 53}, 10871 (1996).

\bibitem{Expt1} 
B. Su, V.J. Goldman, and J.E. Cunningham, Science {\bf
255}, 313 (1992); Phys. Rev. B {\bf 46}, 9644 (1992).

\bibitem{Expt2} R.C. Ashoori, H.L. Stormer, J.S. Weiner,
L.N. Pfeiffer, K.W. Baldwin, and K.W. West, Phys. Rev. Lett. {\bf 71},
613 (1993); Phys. Rev. Lett. {\bf 68}, 3088 (1992).

\bibitem{Expt3} B. Meurer, D. Heitman, and K. Ploog, Phys. Rev.
Lett. {\bf 68}, 1371 (1992).

\bibitem{Tsui} D.C. Tsui, H.L. Stormer, and A.C. Gossard, Phys. Rev.
Lett. {\bf 48}, 1559 (1982).

\bibitem{Jain} J.K. Jain, Phys. Rev. Lett. {\bf 63}, 199 (1989);
Phys. Rev. B {\bf 41}, 7653 (1990).

\bibitem{Mandal} S.S. Mandal and J.K. Jain, 
Phys. Rev. B {\bf 66}, 155302 (2002).

\bibitem{Comment} It so happens for $N=6$ that for 
one point in each panel of Fig.~\ref{fig:energy}, 
the smallest dimension $D^*$ is obtained with an anomalous value of 
$2p$.  We actually get a slightly better CF energy if we instead choose   
here $2p$ that matches $2p$ for other $L$ in the panel.

\bibitem{classical} V.M. Bedanov and F.M. Peeters, Phys. Rev. B 
{\bf 49}, 2667 (1994);
F. Bolton and U. R\"ossler, Superlatt. Microstruct. {\bf 13}, 139 (1993);

\bibitem{melting} V. A. Schweigert and F.M. Peeters, Phys. Rev. B {\bf 51},
7700 (1995); A. V. Filinov, M. Bonitz, and Yu. E. Lozovik, Phys. Rev. 
Lett. {\bf 86}, 3851 (2001).

\bibitem{Fertig} R. Narevich, G. Murthy, and H. A. Fertig,
Phys. Rev. B {\bf 64}, 245326 (2001);
H. Yi and H. A. Fertig, Phys. Rev. B {\bf 58}, 4019 (1998). 

\bibitem{Peterson} S.S. Mandal, M.R. Peterson, and J.K. Jain, 
Phys. Rev. Lett. {\bf 90}, 106403 (2003).

\end{thebibliography}
\end{document}